\documentclass[11pt]{article} 

\usepackage{geometry} 
\title{Quantization of Black Holes entropy and its cosmological consequences}
\author{{G. Cristofano${ }^{1,2}$, G. Maiella${ }^{1,2}$ and C. Stornaiolo${ }^{1}$}\\
 {\em $~^{1}$ Istituto Nazionale di Fisica Nucleare,}{\em Sezione di Napoli,}\\
 {\em  Complesso Universitario di Monte S. Angelo}
 {\em Edificio 6 }\\ {\em via Cinthia, 45 -- 80126 Napoli}\\
{\em $~^{2}$ Dipartimento di Scienze Fisiche,}\\ 
{\em Universit\`{a} ``Federico II'' di Napoli,}\\
 {\em  Complesso Universitario di Monte S. Angelo}
 {\em Edificio 6  }\\ 
 {\em via Cinthia, 45 -- 80126  Napoli}\\ }
\date{ }

 \usepackage{txfonts}

\begin{document}
\maketitle
\abstract{Starting from a quantization relation for primordial extremal black holes with electric and magnetic charges, it is shown that their entropy is quantized. Furthermore the energy levels spacing for such black holes is derived as a function of the  level number $n$, appearing in the quantization relation. Some interesting cosmological consequences are presented for small values of $n$. By producing a mismatch between the mass and the charge  the black hole temperature is derived and its behavior investigated. Finally extending the quantum relation to Schwarzschild black holes  their temperature is found to be in  agreement with the Hawking temperature and a simple interpretation of the microscopic degrees of freedom of  the black holes is given.}
\section{Introduction}
Recently \cite{Capozziello:2010jx}  it has been shown that primordial black holes masses obey a fundamental quantization condition which basically stems from the Dirac consistency relation for the wave function describing the quantum state of electrically and magnetically charged black holes. What is really far reaching is that the masses of astrophysical self-gravitating structures, as galaxies, clusters of galaxies and superclusters and the same universe mass are correctly described by such a quantization condition, so reveling a strict  connection between quantum fluctuations, at work at the very beginning of the universe, and the large scale astrophysical structures observed at   present in the universe.

Furthermore such astrophysical structures lie on  a straight line in a plot of angular momentum $J=n \hbar$ of the structure versus  its squared mass  $m^{2}$ with a universal slope $G/c$. More recently it has been shown that as an outcome of the fundamental quantum relation, also their physical scales are fully derived in terms of the Compton wavelength of its basic constituent, that is the proton \cite{Capozziello:2011rn}. 

In this letter  starting from a numerical coincidence, we derive an analytical expression for the quantum level number  $n$, which allows for an interesting interpretation, More precisely the quantum level $n$ associated to the mass squared  $m^2$  turns out to describe the number of degrees of freedom of the astrophysical structures, that is its entropy, which then appears to be quantized. The concept of energy level spacing is then introduced for black holes as a natural consequence of the quantization relation (\ref{10 }). Also its far reaching cosmological implications are presented, by considering the splitting of the lowest "energy" states.  Furthermore introducing a mismatch $\Delta m$ between the mass and the charge of the black hole its temperature is derived and analyzed in the two interesting limits $\Delta m\gg m$  and $\Delta m \ll m$. Also the time lapse for the excited state to decay to the stationary one is estimated and it is found to go to infinity for $\Delta m$ going to zero . Finally extending the quantum relation to Schwarzschild black holes and using the connection between the quantum level number $n$ and  the entropy of the black hole the first law of thermodynamics for a black hole  is enforced  and its temperature derived, finding full agreement with the Hawking temperature.  More interestingly such an approach allows for a microscopic description of the internal degrees of freedom of the black hole.

The letter is organized as follows, in sect.\ref{s2}  we recall the basic quantization relation emphasizing the numerical coincidence of the quantum level number  $n$ and the maximal entropy for the universe as a black hole. In sect. \ref{s3} we derive the analytical expression for $n$ in terms of the entropy of the astrophysical structure. In sect.   \ref{sec4}  the energy level spacing for a quantized black hole is obtained and its cosmological consequences presented. In sect. \ref{sec5} the   temperature of the perturbed extremal black hole is derived and its  behavior  discussed.  In sect. \ref{sec6}  the thermodynamic properties of Schwarzschild black holes are discussed and the Hawking temperature derived together with the interpretation of its internal degrees of freedom. In sect. \ref{s6} some comments and conclusions are presented.

\section {Black holes quantization}\label{s2}

Let us start from the Reissner-Nordstrom type metric for  a black hole of mass $m $ with electric charge $Q_e$ and magnetic charge $Q_m$,
\begin{equation}\label{1}
ds^2=\left( 1-\frac{r_s}{r}+\frac{r^2_{Q_e}+r^2_{Q_m}}{r^2}\right)c^2dt^2 - \left( 1-\frac{r_s}{r}+\frac{r^2_{Q_e}+r^2_{Q_m}}{r^2}\right)^{-1}dr^2-r^2d\Omega^2
\end{equation}
with
\begin{equation}
\label{ 2}
r_s=2\frac{Gm}{c^2}\ \
\end{equation}
the Schwarzschild radius and  the other significant lengths
\begin{equation}
\label{3 }
r^2_{Q_e}=\frac{Q^2_e G}{4\pi \epsilon_0 c^4}\ \ \  \mathrm{and }\  \ \ r^2_{Q_m}=\mu_0\frac{Q^2_m G}{4\pi c^4} ,
\end{equation}
where $Q_{e}$ and $Q_{m}$ indicate the electric and magnetic charges of the black hole.  Furthermore, by imposing the condition  $g_{00}=0$ we get for the   event horizons the following solutions,
\begin{equation}
\label{ 4}
r_\pm = \frac{r_s \pm \sqrt{r^2_s -4(r^2_{Q_e} + r^2_{Q_m} )}}{2}.
\end{equation}
For extremal black holes $r_+=r_-$ implying
\begin{equation}
\label{ 5}
Gm^2=\frac{Q^2_e}{4\pi \epsilon_0 }+\mu_0\frac{Q^2_m }{4\pi }=q^{2}_{e}+q^{2}_{m}
\end{equation}

The preceding relations can be derived from a Lagrangian formulation  of extremal black holes.    Basically there are two approaches to it. A non-supersymmetric   \cite{Goldstein:2009cv} and a supersymmetric one \cite{Ferrara:2008hwa}  in which there are $N$  scalar fields  $\phi_{i}$ coupled to gauge fields with a dilaton-like coupling.   In both cases an effective potential $V_{eff.}(\phi_{i})$ is derived and a so-called attractor mechanism takes place once the following  conditions are satisfied.
$$\partial_{i}V_{eff.}(\phi_{i0})=0$$
where $\phi_{i0}$ are critical field values; the matrix
$$M_{ij}=\frac{1}{2}\partial_{i}\partial_{j}V_{eff.}(\phi_{i0})$$
has positive eigenvalues.

In this context  in the $N=2,4$ supersymmetric theories the above conditions are automatically satisfied. The effective potential $V_{BH}$ for a black hole with electric and magnetic charges in supergravity $D=4$, $N=4$ is
\begin{equation}
\label{ bh1}
V_{BH}(\phi,a, q_{e},q_{m})=e^{2\phi}\left(q_{e}-aq_{m}\right)^{2}+e^{-2\phi}q_{m}^{2} .
\end{equation}
where $\phi$ is the dilaton field, $a$ the axion field and only one component of the electric  and magnetic charges is considered.  We can find a correspondence of this potential in conformal field theory (CFT) where the effective potential is obtained in the CFT description of a quantum Hall fluid,
\begin{equation}
\label{bh2}
V_{eff.}^{CFT}=R^{2}_{c}\left(Q_{e}-\frac{\theta}{2\pi} Q_{m}\right)^{2}+\frac{1}{R^{2}_{c}}Q^{2}_{m}
\end{equation}
where $R^{2}_{c}$  is the compactification radius  of the scalar Fubini field \cite{Fubini:1990bw} and $\theta$ is the theta parameter introduced by 't Hooft \cite{'tHooft:1977hy}
\cite{Cardy:1981qy}.

There is a striking resemblance between the two potentials (\ref{ bh1}) and (\ref{bh2}), which shows that the black hole physics can be described in terms of a CFT (see also \cite{Maiella:2006hr}) and 
  the following identifications
\begin{equation}
\label{ bh3}
R^{2}=e^{2\phi}, \ \ \ \ \ \ \   \frac{\theta}{2\pi}=a.
\end{equation}
can be made.

We must notice that having considered just one  electric and magnetic charge component in eq.(\ref{ bh1}), the axion field $a$  and then $\theta$ in eq. (\ref{bh2}) has to be taken equal to zero for stability reasons (see ref. \cite{Ferrara:2008hwa}). Then the effective potential we will consider from now on becomes the following, 
\begin{equation}
\label{ 6}
V_{BH}(\phi,a, q_{e},q_{m})= e^{2\phi}q^2_e  + e^{-2\phi}q^2_m.
\end{equation}
By taking the field $\phi$ constant  we can determine its value by the stability condition $\partial V_{BH} /\partial \phi=0$ keeping fixed the charges, obtaining 
\begin{equation}
\label{7 }
e^{2\phi_H}=\frac{q_m}{q_e}.
\end{equation}
Such a procedure is equivalent to the requirement of the double extremality condition used in \cite{Ferrara:2008hwa}. As a consequence the following relation    
\begin{equation}
\label{ 6b}
 V_{BH}(\phi_{H},a, q_{e},q_{m})\equiv e^{2\phi_{H}}q^2_e  + e^{-2\phi_{H}}q^2_m=Gm^2,
\end{equation}
gets satisfied (see in particular  equation (5.13) in \cite{Ferrara:2008hwa}). Notice that the black hole mass depends only on the strength of the charges, in fact substituting the value at the horizon given by eq (\ref{7 }) into  eq. (\ref{ 6b}) we obtain the following relation
\begin{equation}
\label{ 8}
Gm^2=2q_eq_m.
\end{equation}
By  employing Dirac consistency relation \cite{Dirac:1931kp} for a quantum description for such extremal black holes,
\begin{equation}
\label{9 }
2q_eq_m= n\hbar c,
\end{equation}
we get finally 
\begin{equation}
\label{10 }
Gm^2=n\hbar c\ \ \ \ \ \ \ n=\mathrm{integer}>0.
\end{equation}
For the lowest allowed mass for an extremal quantum black hole we get
\begin{equation}
\label{11 }
m=\sqrt{\frac{\hbar c}{G}}=m_{Planck}
\end{equation}
That is, according to the condition $g_{00}=0$,  we can suggest that at its very beginning $t\approx 10^{-43}\, s$ and at the Planck temperature $T_{Planck}=10^{32}\ \textrm{K}$ charged black holes were forming, thanks to the balance between the attractive gravitational force and the repulsive electric and magnetic forces.

Astrophysical and cosmological observations suggest that the previous quantum relation (\ref{10 })  found for dilatonic charged black holes  in their extremal regime correctly describes also  the astrophysical and cosmological structures at any scale \cite{Capozziello:2010jx}
 \cite{Capozziello:2011rn}, and suggests that it can be applied to   black holes of any class. In section \ref{sec6} another argument in favor of this last statement will be presented,  by studying their thermodynamical properties.  In the following  sections, starting from the above quantization relation, the black hole entropy will be expressed in terms of the quantum number $n$,  the energy levels for the allowed black holes  and its temperature derived. 

\section{Quantization of black holes entropy}\label{s3}

A further interesting comment can be made regarding the  generality of the  quantization relation (\ref{10 }), that is there is no remnant of the charges of the allowed quantum black holes, instead there appears its angular momentum $n\hbar$ as a quantized quantity.
That suggests   the quantum relation (\ref{10 }) to be very general and indeed its validity has been proven to extend to astrophysical structures as galaxies, clusters of galaxies, superclusters and the whole universe (scaling hypothesis) \cite{Capozziello:2010jx}\cite{Capozziello:2011rn}. That is the quantum relation (\ref{10 }) represents a basic, conceptual link, relating cosmological structures to quantum fluctuations at  primordial epochs. It  extends up to the whole (observable) universe  
\begin{equation}
\label{12 }
Gm^2_{U} =n_{U}\hbar c.
\end{equation}
In fact by using $m_{U}= N_p m_p$ where  $m_p=10^{-31}\ \mathrm{g} $  is the proton mass and $N_p \approx 10^{80}$ is the number of protons in the observed universe we obtain  $n_U\approx (10^{60})^2=10^{120}$,  just what it is expected for the universe total action \cite{Capozziello:2010jx}.

This huge number seems to be related with the  maximal  possible entropy of the universe, as if it were  a black hole as a whole, with radius equal to the event horizon,  as suggested by the holographic principle \cite{Bousso:2002ju} for a black hole 
\begin{displaymath}
\frac{S_{max}}{k_B}\sim 10^{120}
\end{displaymath}
where $k_B$ is the  Boltzmann constant.

We now show that such a  coincidence is  not merely numerical. Indeed, assuming the generality of  equation (\ref{10 }),  the quantization of the black hole entropy immediately follows. In fact the explicit  expression for the quantum level number $n$  appearing in reference  \cite{Capozziello:2010jx}
\begin{equation}
\label{ 13c}
n=\frac{1}{2}\frac{r_s}{\lambdaslash_{Compt.}}
\end{equation}
is, apart from a numerical factor,   nothing else but the Bekenstein-Hawking entropy for an extremal black hole. To this end the following relation can be easily verified,
\begin{equation}
\label{14 }
\frac{S}{k_B}=\frac{1}{4}\frac{4\pi r_+ ^{2} }{\ell_P^2}=\frac{1}{4}\frac{4\pi }{\ell_P^2}\left(\frac{r_s}{2}\right)^{2}=\pi\frac{G^2 m^2}{c^4}\frac{c^3}{\hbar G}=\pi\frac{1}{2}\frac{r_s}{ \lambdaslash_{Compt.}}= \pi n
\end{equation}
where $\displaystyle \lambdabar_{Compt.}=\frac{\hbar}{mc}$ is the   reduced Compton wavelength corresponding to the mass $m$.
%

The above relation implies  that the entropy is quantized, due to the appearance of the level quantum number $n$ in the last step. The other interesting result is that the black hole entropy is expressed also by the ratio between the Schwarzschild radius and the  Compton wavelength associated to the black hole mass.

A consequence of these two statements is that  the Schwarzschild radius is an integer number of times the Compton wavelength which then appears to be a fundamental length in black hole physics.
\section{Energy level spacing for a quantized  black hole}\label{sec4}

It is interesting to show that the quantum relation (\ref{10 }) allows us to introduce  the concept of energy level spacing between two energy levels for a quantized black hole
\begin{equation}
\label{21}
\frac{\Delta E}{E_{P}}= \frac{(m_{n+1} - m_n) c^2}{m_P c^2},
\end{equation}
where $E_P$ is the Planck energy.

We can evaluate $ (m_{n+1} - m_n)  $  from equation (\ref{10 }) by deriving the following quantity
\begin{equation}
\label{22}
\frac{ m_{n+1}^2 - m_n^2 }{m_n^2}=\frac{ (m_{n+1} - m_n)  (m_{n+1} +m_n) }{m_n^2}=\frac{1}{n}.
\end{equation}
For $n>>1$,  $  m_{n+1}\approx  m_n $ and we can write safely
 \begin{equation}
\label{23}
 (m_{n+1} - m_n) = \frac{1}{2n}m_n=\frac{1}{2}\frac{c\hbar}{Gm_n^2}m_n=\frac{1}{2}\frac{m_P^2}{m_n}
\end{equation}
and
\begin{equation}
\label{24}
\frac{\Delta E}{E_P}= \frac{1}{2}\frac{m_P}{m_n}\propto \frac{1}{\sqrt{n}}.
\end{equation}
Similar results can be found in a different context in  \cite{Hernandez:2005sf}.  We can see that  the larger is the mass, the smaller is the energy needed for a transition from one mass to a  higher  one. That is a black hole of larger mass can ``swallow'' almost anything more  easily than  black holes  of smaller masses \cite{Bekenstein:1997bt} . If we evaluate   relation (\ref{22})  for the lowest value $n=1$ a far reaching cosmological result  can be obtained. In fact for $n=1$ we obtain  the following relation, 
\begin{equation}
\label{24b }
 \frac{Gm_{2}^{2}-Gm_{planck}^{2}}{Gm_{planck}^{2}}=\frac{2\hbar c-\hbar c}{\hbar c}=1.
  \end{equation}
 That is 
 \begin{equation}
\label{24c }
 \frac{(m_{2}-m_{planck})(m_{2}+m_{planck})}{m_{planck}^{2}}=1
\end{equation}
 and finally
\begin{equation}
\label{2d }
 \frac{\Delta E}{E_{Pl}}=\frac{(m_{2}-m_{pl})}{m^{2}_{Pl}}=\frac{m_{pl}}{m_{2}-m_{pl}} \simeq \frac{1}{2}
\end{equation}
 That is the energy level splitting between the ''first excited  level" and the ''ground state level" is of the order of the Planck energy.  That is below the temperature
\begin{equation}
\label{dandc }
T<\frac{1}{2}T_{P},
\end{equation}
due to the expansion of the universe,  gravitons \cite{weinberg} do not have enough energy to be absorbed from the primordial quantum lowest mass black holes due to quantization and consequently they decouple from matter.

In our opinion these high energetic gravitons would travel isotropically in all directions and fill now all the space within the cosmological structures. It would be  interesting to detect their "relics" now and trace back a picture of the universe at   Planck time. Further details on that will be reported elsewhere.

A further question is  now  in order,  together with its  decoupling from  matter are we allowed to say that at   $T_{planck}/2$ gravity decouples from the other still unified forces, i.e. the strong, the weak and the electromagnetic forces?


\section{Heuristic derivation of black hole temperature and its quantization}\label{sec5}

Let us start  noticing that  the quantization relation given in eq. (\ref{10 }) was derived just under  the  assumption that  black holes, with electric and magnetic charges, were extremal that is  for  $r_{+}=r_{-}$ which is equivalent to   temperature $T=0$ for them.
Furthermore  by changing $n$, through higher and higher positive integer values the energy levels described in eq.(\ref{10 }) correspond to higher and higher mass black holes, which do not radiate due to its zero temperature; that is such states appear to be stable or stationary.

By introducing   a mismatch between the mass and the charge, the temperature will be different from zero according to  
\begin{equation}\label{}
    T=\frac{\hbar c }{2\pi   k_{B} }  \frac{r_{+}-r_{-}}{2r^{2}_{+}}
\end{equation}
That is explicitly
\begin{equation}\label{}
 T=\frac{\hbar  c}{ \pi   k_{B}  } \frac{\sqrt{r^{2}_{s}-4(r^{2}_{e}+r^{2}_{m})}}{\left( r^{2}_{s} +2r_{s} \sqrt{r^{2}_{s}-4(r^{2}_{e}+r^{2}_{m})} + r^{2}_{s}-4(r^{2}_{e}+r^{2}_{m})\right) }
 \end{equation}
 If we perturb the extremal black hole by changing only its mass according to  $$m\to m+\Delta m,$$  we obtain for the temperature 
the following expression,  
\begin{equation}\label{}
 T=\frac{\hbar  c^{3} }{2 \pi  G  k_{B} }\frac{\sqrt{ (m+\Delta m)^{2} -Q^{2}}}{\left( (m+\Delta m)^{2} +2(m+\Delta m)\sqrt{(m+\Delta m)^{2}- Q^{2}} + (m+\Delta m)^{2}-Q^{2}\right) }
\end{equation}
where we defined 
$$Q^{2}=\frac{4}{G}(r^{2}_{e}+r^{2}_{m})$$
After imposing the condition  $m^{2}=Q^{2}$ we get,
\begin{equation}\label{tx}
 T=\frac{\hbar c^{3}}{ 8\pi G k_{B}m} \frac{4\sqrt{x ^{2} +2x }}{\left( 1 +  2x ^{2} +4x +2(1+x)\sqrt{x ^{2} +2x}\right) }
\end{equation}
where $x=\Delta m/m$.
It is instructive to consider the two extreme cases $x \gg 1$ and $x \ll 1$.

For  $x \gg 1$, keeping only the leading terms, one easily gets 

\begin{equation}\label{}
 T=\frac{\hbar c^{3}}{ 2\pi G k_{B}m} \frac{\sqrt{x ^{2}}}{\left(2x ^{2}+2x\sqrt{x^{2}}\right) }
 =\frac{\hbar c^{3}}{ 8\pi G k_{B}m}\frac{1}{x},
\end{equation}

which   is the temperature of a Schwarzschild black hole of mass $\Delta m$.

For the  $x \ll 1$ case, after neglecting smaller terms, one gets

\begin{equation}\label{bh10}
 T= \frac{\hbar c ^{3}}{ 8\pi G k_{B}m}\cdot 4 \sqrt{2x}
  \end{equation}
or
 \begin{equation}\label{bh11}
 T=\frac{\hbar c ^{3}}{ 2\pi G k_{B}} \frac{\sqrt{2m\Delta m }}{   m^{2}  }=
 \frac{c^{2}}{2\pi k_{B}} \left(\frac{\hbar c}{G} \right)^{1/4}\frac{\sqrt{2\Delta m}}{n^{3/4}}
\end{equation}
where use has been made of the quantization condition given in eq. (\ref{10 }). Notice that in eq. (\ref{bh10})   the deviation of the black hole temperature from the temperature of a Schwarzschild black hole of equal mass is explicitly shown, while in eq. (\ref{bh11}) the relation of the temperature with the quantum number $n$ has been evidenced.

Let us now determine the time required for a perturbed charged black hole to reach  the  extremal state.

One must solve the differential equation
\begin{equation}
\label{decay1}
-\frac{dE}{dt}=P
\end{equation}
where
\begin{equation}
\label{decay2}
P=A\epsilon\sigma T^{4}_{H}
\end{equation}
is the radiation power of a black hole, being $A$ the black hole surface area, $\sigma=\pi^{2}k_{B}/ 60 \hbar^{3} c^{2}$  the Stefan-Boltzmann constant. Assuming that the spectrum of the black hole is a black body spectrum ($\epsilon=1$), get 
\begin{equation}
\label{decay3}
P=\frac{\hbar c^{6}}{15360 \pi G^{2} m^{2}} \frac{64(x ^{2} +2x )^{2}}{\left( 1 +  2x ^{2} +4x +2(1+x)\sqrt{x ^{2} +2x}\right)^{3} }
\end{equation}
and  eq. (\ref{decay1}) becomes
 \begin{equation}
\label{decay4}
-\frac{dE}{dt}\equiv -mc^{2}\frac{dx}{dt}=\frac{\hbar c^{6}}{15360 \pi G^{2} m^{2}} \frac{64(x ^{2} +2x )^{2}}{\left( 1 +  2x ^{2} +4x +2(1+x)\sqrt{x ^{2} +2x}\right)^{3} }
\end{equation}
Solving  this equation it is easy to see that , 
\begin{equation}
\label{decay5}
\int_{x_{0}}^{0}  \frac{\left( 1 +  2x ^{2} +4x +2(1+x)\sqrt{x ^{2} +2x}\right)^{3} }{64(x ^{2} +2x )^{2}}dx=-\frac{\hbar c^{6}}{15360 \pi G^{2} m^{2}} (t-t_{0}).
\end{equation}
Notice that the integral on the lefthand side diverges  in agreement  with the third law of thermodynamics which states that the zero temperature cannot be reached in a finite time. That is equivalent to say that the derivative $dx/dt$ goes to zero very fast and consequently  the emission of the Hawking radiation slows down,   see eq. (\ref{decay4}).

\section{Extending the quantum relation to more general black holes}\label{sec6}

We are now ready to enforce the statement that the quantum relation obtained for extremal black holes is indeed valid for black holes of any class. Here we will consider just the Schwarzschild ones. To such an extent we will derive the temperature of a black hole, just using  relation (\ref{10 }) and the established connection between its entropy and the level quantum number $n$
\begin{equation}
\label{14b }
\frac{S}{k_B}=\frac{1}{4}\frac{4\pi r_s ^{2} }{\ell_P^2}= 2\pi\frac{r_s}{ \lambdaslash_{Compt.}}=4 \pi n,
\end{equation}
which is an extension of  equation (\ref{14 }) to Schwarzschild black holes.
It is interesting to notice that by using these two relations it is possible to enforce the first principle of thermodynamics and derive from it the black hole temperature $T_{BH}$, which  is found to agree with the Hawking temperature, but in such a new context it is seen naturally to be a quantized quantity.

First let us start with the assumption that  the quantum relation
\begin{equation}
\label{tbh1b }
Gm^{2}=n\hbar c
\end{equation}
extends to Schwarzschild black holes.  Indeed  we recall that several authors have tried to introduce a quantization for them, through analogous methods, like quantization of black hole horizon or Ehrenfest quantization of their action,  see refs. \cite{Hernandez:2005sf}  \cite{Bekenstein:1997bt} \cite{Bekenstein:1995ju} among others.  

By differentiating  left and right hand in the previous relation we obtain,
\begin{equation}
\label{TBH2}
2Gmdm=\hbar c dn .
\end{equation}
Furthermore using
%
\begin{equation}
\label{tbh4 }
\frac{dS}{k_{B}} =4\pi dn
\end{equation}
we get
\begin{equation}
\label{tbh5}
2Gmdm=\frac{1}{4\pi}\frac{\hbar c}{k_{B}}dS.
\end{equation}
Dividing by the factor $2Gm$ and multiplying by $c^{2}$ get
\begin{equation}
\label{tbh7}
dm c^{2}=\frac{1}{8\pi}\frac{\hbar c^{3}}{k_{B}G m}dS,
\end{equation}
which states the first law of thermodynamics for a Schwarzschild black hole,
\begin{equation}
\label{tbh8}
dU= T_{BH} dS
\end{equation}
with
\begin{equation}
\label{tbh9}
T_{BH}=\frac{1}{8\pi} \frac{\hbar c^{3}}{k_{B}G m}=\frac{1}{8\pi k_{B}} \frac{mc^{2}}{n}
\end{equation}
which agrees with the Hawking temperature. Moreover, as it appears in the last step of eq. (\ref{tbh9}), the black hole  temperature $ T_{BH}$ is quantized as a consequence of the quantum relation (\ref{tbh1b }).
%
%
%
%
%

We are now ready to give a microscopic description of the internal degrees of freedom of a black hole, and we will   discover that it naturally arises a proposal of other different scenarios, possibly at work.

Let us start from the simplest case of  lowest black hole mass   corresponding to $m=m_{P}$.  Then eq. (\ref{tbh9}) specialized for the case $n=1$ allows us to give a direct interpretation of the microscopic internal degrees of freedom of the elementary black hole. In fact it is straightforward to get the following relation.
\begin{equation}
\label{new1}
m_{P}c^{2}= 4\pi\cdot 2 k_{B}T_{P}
\end{equation}
the factor $2 k_{B}T_{P}$ is the average energy associated to a two-dimensional oscillator in thermal equilibrium at Planck temperature $T_{P}$. Notice  that  $4\pi$ is the same incommensurability factor appearing in  eq. (\ref{14b }) in defining the entropy as the ratio between  the black hole area and the elementary area $\ell_{p}^{2}$.  

We can now go back to eq.  (\ref{tbh9}) to get a more general result regarding the total energy of a black hole with a generic mass obtaining analogously 
\begin{equation}
\label{new2}
mc^{2}=4\pi n \cdot 2 k_{B}T_{BH}
\end{equation}
and  the factor   $2 k_{B}T_{BH}$  is the average energy of each two-dimensional harmonic oscillator  at thermal equilibrium at temperature $T_{BH}$ and consequently $n$ can be  interpreted as the number of oscillators filling the horizon in a two-dimensional array.  

 In addition to the above description of the internal degrees of freedom, we present here two more possible scenarios, among others,

%

1) $\displaystyle mc^{2}=4\pi  n_{F1/2}\times \frac{1}{2}k_{B}T_{BH}\times 2$ (spin degrees of freedom) $\times 2$ (translational degrees of freedom),

2)  $\displaystyle mc^{2}= 4 \pi n_{F3/2 }\times \frac{3}{2}k_{B}T_{BH}\times 4$ (spin degrees of freedom).

where $  n_{F1/2}$ and $  n_{F3/2}$ are  the   number of fermions of spin $s=1/2$ and   $s=3/2$ respectively. We must comment that all such scenarios stem from the quantum relation (\ref{tbh1b }) together with the interesting interpretation of the level number $n$, appearing in it, and in the black hole entropy given by eq. (\ref{14b }). 

\section{Discussion and conclusions}\label{s6}
In this paper it has been presented a  quantization of a black hole entropy from first principles. More precisely it has been shown that it arises from the quantization relation given in (\ref{10 }), which stems from the Dirac relation  for   electrically and magnetically charged black holes. 

To obtain this result we used a strict similarity between the effective potentials for extremal black holes and conformal field  theory. The link between black holes and conformal field theory was already put in evidence previously in   \cite{Maiella:2006hr},  where  the properties of the horizon of the BTZ
black holes in ADS$_{3}$ were  described in terms of an effective unitary CFT$_{2}$, with central
charge c = 1,  realized in terms of the Fubini-Veneziano vertex operators \cite{Fubini:1990bw}.

We must observe that before this presentation the constant $\hbar$ appearing in the expression of the  black hole entropy  was just  associated to  an ``ad hoc'' introduction of a minimal Planck area, in the definition of black hole ``degrees of freedom'' from its horizon. Furthermore it has been shown the existence of energy levels for the black holes and their spacing as a decreasing function of the energy level.

By considering the lowest lying energy levels it has been evidenced that gravitons decouple as soon as the temperature decreases, due to  expansion,  below  $T_{Planck}/2$ and their "relics" could be experimentally detected.  

Also  keeping the black hole's charge fixed and increasing its mass the emission temperature has been shown to be equal to the Hawking temperature of a neutral black hole, with the same mass, multiplied by a factor  $x=\Delta m/m$.   It is easy to see  from eq. ( \ref{tx}) that for increasing   mass, the temperature first increases, reaches a maximum and    finally,  as in the case of  a Schwarzschild black hole, decreases.  The variation of mass due to the Hawking radiation is continuous, but differently from the Schwarzschild black hole it takes an infinite time to   evaporate all the excess mass. The total evaporation of the black hole could be obtained by bombarding the black hole with charges of the opposite sign until it becomes totally neutral. 

Moreover, extending the quantum relation to Schwarzschild black holes,  its temperature $T_{BH} $ has been derived in a  natural way, within our framework, showing agreement with the Hawking temperature and  a simple interpretation of the microscopic degrees of freedom of the black hole provided. Even though interesting ideas have been presented here regarding black hole entropy quantization, its energy levels, its temperature quantization and a picture of its microstates, much more has to be done towards the understanding of problems like their interaction and/or the role of primordial black hole in the missing mass problem in the universe.
 
 Finally, being the universe at Planck time a quantum  mechanical system and knowing that a phase transition is at work at  $T=T_{Planck}$ with the consequent formation of the lowest mass primordial black holes ($m=m_{Planck}$) one is very tempted to describe the state wave function of the universe at Planck time in terms of a quantum fluid of highly correlated lowest mass black holes, but that will be given elsewhere.


\begin{thebibliography}{99}

\bibitem{Capozziello:2010jx}
  S.~Capozziello, G.~Cristofano and M.~De Laurentis,
  Eur.\ Phys.\ J.\  C {\bf 69} (2010) 293
  [arXiv:1005.2891 [gr-qc]].
 
\bibitem{Capozziello:2011rn}
  S.~Capozziello, G.~Cristofano and M.~De Laurentis,
   Mod.Phys.Lett. A26 (2011) 2549-2558 
  arXiv:1110.1175 [gr-qc].

\bibitem{Goldstein:2009cv}
  K.~Goldstein, S.~Kachru, S.~Prakash and S.~P.~Trivedi,
  JHEP {\bf 1008} (2010) 078
  [arXiv:0911.3586 [hep-th]].
  
\bibitem{Ferrara:2008hwa}
  S.~Ferrara, K.~Hayakawa and A.~Marrani,
  Fortsch.\ Phys.\  {\bf 56} (2008) 993
  [arXiv:0805.2498 [hep-th]].
  
\bibitem{Fubini:1990bw}
  S.~Fubini,
  Mod.\ Phys.\ Lett.\ A {\bf 6} (1991) 347.
  
\bibitem{'tHooft:1977hy}
  G.~'t Hooft,
  Nucl.\ Phys.\ B {\bf 138} (1978) 1.
  
  
\bibitem{Cardy:1981qy}
  J.~L.~Cardy and E.~Rabinovici,
  Nucl.\ Phys.\ B {\bf 205} (1982) 1.
  
  
\bibitem{Maiella:2006hr}
  G.~Maiella and C.~Stornaiolo,
  Int.\ J.\ Mod.\ Phys.\ A {\bf 22} (2007) 3429
  [hep-th/0611194].

\bibitem{Dirac:1931kp}
  P.~A.~M.~Dirac,
  Proc.\ Roy.\ Soc.\ Lond.\ A {\bf 133} (1931) 60.


\bibitem{Hernandez:2005sf}
  X.~Hernandez, C.~S.~Lopez-Monsalvo, S.~Mendoza and R.~A.~Sussman,
  Rev.\ Mex.\ Fis.\  {\bf 52} (2006) 515
  [gr-qc/0507022].


\bibitem{Bekenstein:1997bt}
  J.~D.~Bekenstein,
  ``Quantum black holes as atoms,'' in Proceedings of
the Eighth Marcel Grossman Meeting on General Relativity, eds. Pirani, T. and Ruffini, R.
(World Scientific).
  gr-qc/9710076.
  
\bibitem{Bekenstein:1995ju}
  J.~D.~Bekenstein and V.~F.~Mukhanov,
  Phys.\ Lett.\ B {\bf 360} (1995) 7
  [gr-qc/9505012].
  
\bibitem{Bousso:2002ju}
  R.~Bousso,
  Rev.\ Mod.\ Phys.\  {\bf 74} (2002) 825
  [hep-th/0203101].


\bibitem{weinberg}
S. Weinberg "Gravitation and Cosmology: principles and applications of the general theory of Relativity''  John Wiley\& Sons (1972)


\end{thebibliography}
\end{document}